# Highly Sensitive MoS$_2$ Photodetectors with Graphene Contacts.

*Peize Han\*[†], Luke St. Marie[†], Qing X. Wang[§], Nicholas Quirk[†], Abdel El Fatimy[†], Masahiro Ishigami[§] and Paola Barbara\*[†]*

[†]Department of Physics, Georgetown University, Washington, DC 20057, USA.

[§]Department of Physics and nanoscience Technology Center, University of Central Florida, Orlando, FL 32816, USA.

*e-mail: ph523@georgetown.edu; paola.barbara@georgetown.edu

ABSTRACT: Two-dimensional materials such as graphene and transition metal dichalcogenides (TMDs) are ideal candidates to create ultra-thin electronics suitable for flexible substrates. Although optoelectronic devices based on TMDs have demonstrated remarkable performance, scalability is still a significant issue. Most devices are created using techniques that are not suitable for mass production, such as mechanical exfoliation of monolayer flakes and patterning by electron-beam lithography. Here we show that large-area $MoS_2$ grown by chemical vapor deposition and patterned by photolithography yields highly sensitive photodetectors, with record shot-noise-limited detectivities of $8.7 \times 10^{14}$ Jones in ambient condition and even higher when sealed with a protective layer. These detectivity values are higher than the highest values reported for photodetectors based on exfoliated $MoS_2$. We study $MoS_2$ devices with gold electrodes and graphene electrodes. The devices with graphene electrodes have a tunable band alignment and are especially attractive for scalable ultra-thin flexible optoelectronics.

MANUSCRIPT TEXT: The variety of electronic properties of two-dimensional materials, ranging from gapless quasi-metals like graphene[1] and $WTe_2$[2], to semiconductors with large direct bandgaps like $WS_2$[3] and $MoS_2$,[4, 5] to insulators like hexagonal boron nitride[6], are a key aspect for the development of multifunctional devices made with different monolayers.[7-9] Graphene/TMD heterostructures are particularly suited for optoelectronic applications because, notwithstanding their monolayer thickness, they exhibit exceptional sunlight absorption. For example, graphene/$MoS_2$ solar cells have been predicted to yield power conversion efficiencies 1-3 orders of magnitude higher than the best existing commercial solar cells reduced to the same thickness[10] and some optoelectronic devices based on graphene/TMD stacks have already demonstrated

superior performance.[11-14] Moreover, since the Fermi energy of graphene can be shifted by a gate voltage or doping, graphene can be used as a contact to semiconducting TMDs yielding tunable Schottky barriers.[15-17] However, these heterostructure devices are most commonly fabricated by stacking flakes obtained from mechanical exfoliation of bulk samples and by patterning device designs with e-beam lithography.[18] These procedures are time-consuming but they are typically preferred because exfoliated samples provide high-quality materials and e-beam lithography is prone to leave fewer residues on the monolayers than standard lithography.[19] Nevertheless, with recent progress in the growth of high quality, large-area graphene and monolayer TMDs, it is now possible to produce high-quality devices without using mechanical exfoliation. For example, large-area materials grown by chemical vapor deposition (CVD) were recently successfully used to fabricate good-quality graphene/$MoS_2$ field-effect transistors [20] and flexible photodetectors.[21] However, in most cases the performance of devices from exfoliated material cannot be matched by devices based on CVD-grown materials. This is due to both the low crystallinity of large-area CVD grown TMDs and the transfer process from their growth substrate often affecting their optical properties, due to unwanted doping or to structural changes during transfer.[22-25]

Here we study $MoS_2$ photodetectors fabricated by combining standard photolithography with transfer techniques of CVD-grown monolayers, without degrading the material properties. Our devices yield performance that matches or exceeds the performance of photodetectors based on exfoliated materials, with detectivities up to $8.7 \times 10^{14}$ Jones in ambient conditions and even higher when encapsulated. This figure of merit is at least one order of magnitude higher than the values reported for exfoliated devices.[26] We also compare devices with gold electrodes and devices with graphene electrodes to study the effect of the different band alignment between the electrodes and the $MoS_2$ on the electrical characteristics.

We grow the graphene by CVD on copper films using the process described in ref.[27] Before the growth, the copper is annealed for recrystallization.[28] Details of the growth process are included the Methods section. We transfer the graphene from the copper film to a doped silicon substrate covered with 300 nm of $SiO_2$, using standard polymethyl methacrylate (PMMA) coating and ammonium persulfate (APS) copper etchant at room temperature.[29] We pattern the graphene using two layers of PMMA (200 nm) and a third layer of SU-8 2002 (2 μm), following the procedure described in our previous work [30-32]. Since the photoresist in contact with the graphene is PMMA, the same photoresist typically used for e-beam lithography, this process does not introduce any additional contamination or residues compared to standard e-beam lithography. We note that annealing procedures aimed to substantially reduce the residues left on graphene from processing with PMMA[33] and other photoresists[34] have been studied before. Although we do not use any annealing in the work presented here, a standard annealing step to clean PMMA residues from the patterned graphene can easily be added to our process (see the Supplementary Information).

For the growth of large-area $MoS_2$, we use the seeding promoter perylene-3,4,9,10-tetracarboxylic acid tetrapotassium salt (PTAS) with $MoO_3$ and sulfur powder precursors, similar to previous work.[35, 36] We pattern the $MoS_2$ film on the growth substrate into microribbons using the photolithography process described above. The patterning is followed by $O_2$ reactive ion etching (with a flow of 50 sccm, 60W forward power for 4 minutes) to etch the parts of the film not covered by the PMMA/SU8 photoresist. The layer of PMMA/SU8 can easily be removed from the $MoS_2$ microribbons with acetone and isopropanol.

Next, we transfer the $MoS_2$ from the growth substrate to the device substrate, and do so without damaging it or degrading its optical properties. The transfer procedure is described in the Supplementary Information. Last, we pattern Au contacts to the graphene electrodes. Figure 1 (a) and (b) show a $MoS_2$ microribbon aligned between two graphene electrodes. The Raman spectra corresponding to different points in the heterostructure are in Figure 1 (f), (g) and (h). In the graphene Raman spectrum, shown in Fig. 1(f), the ratio of the 2D to G peak is larger than two, the width of the 2D peak is 29 cm$^{-1}$ and there is no noticeable D peak, indicating no substantial increase of defects due to processing.[37] Figure 1 (h) shows the Raman and photoluminescence spectra of the patterned $MoS_2$ film before and after the patterning, transfer and alignment procedures. The slight shift of the Raman and photoluminescence peaks after transfer has been observed by other groups and explained with the release of strain in the $MoS_2$ (the mismatch of thermal expansion coefficients between the $MoS_2$ and the substrate causes strain in the as-grown $MoS_2$).[38] Although chemical treatments and defects can change the photoluminescence by more than one order of magnitude,[38, 39] the photoluminescence peak is reduced only by about a factor of two, indicating no substantial change in the film properties. The photoluminescence mapping in Figure 1(c) shows the uniformity of the $MoS_2$ film.

To compare the results from our CVD-grown $MoS_2$ to those of studies of photodetectors using exfoliated $MoS_2$, we also fabricated samples with the same geometry as the graphene/$MoS_2$/graphene samples, but with gold electrodes instead of graphene electrodes, since most previous work on high-performance $MoS_2$ photodetectors has used gold electrodes.[26, 40, 41] We patterned the gold electrodes by lift-off using Shipley S1813 photoresist. Figure 1 (e) shows a photoluminescence map of the $MoS_2$ film between the Au electrodes, confirming the continuity and uniformity of the film after transfer.

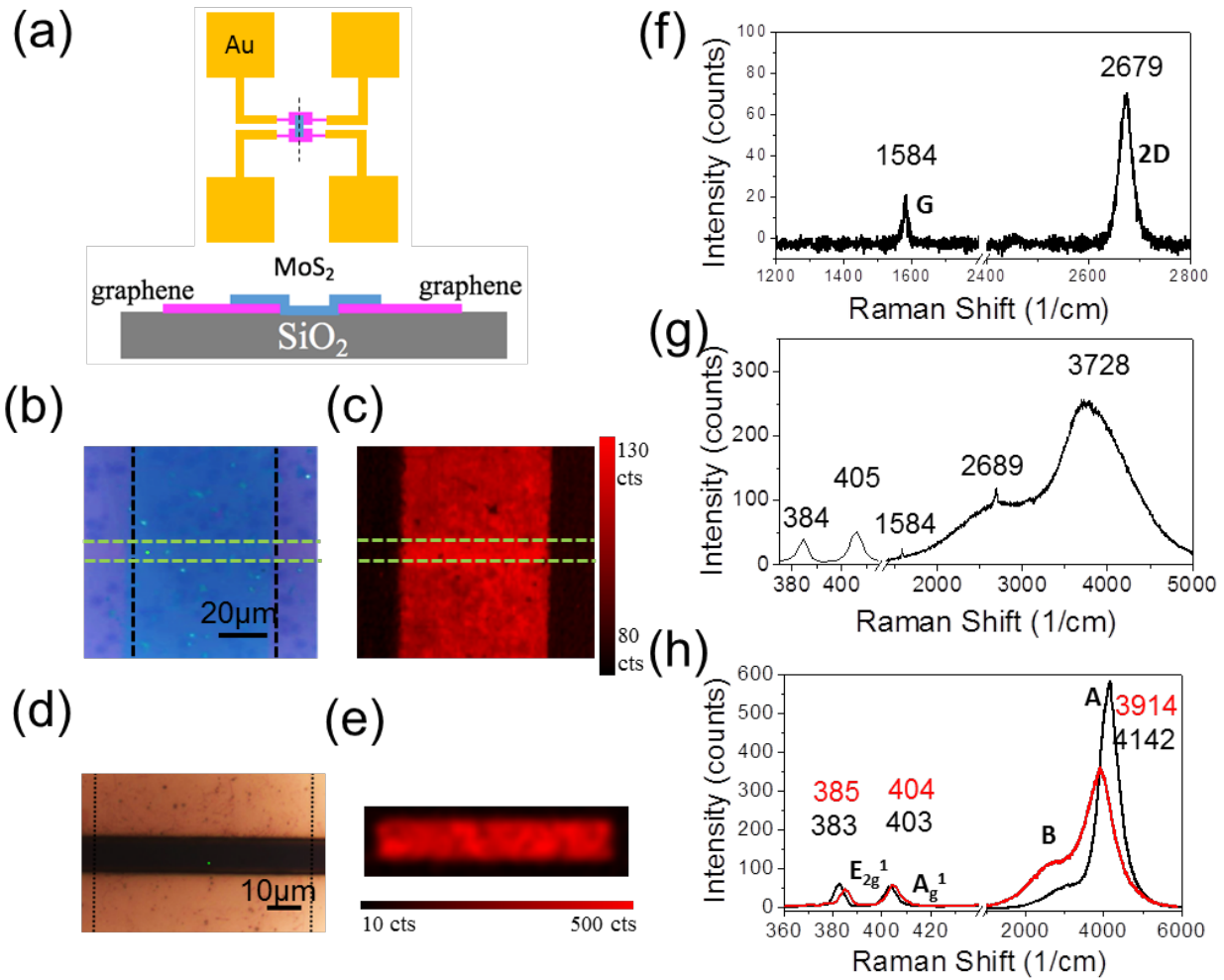

**Figure 1.** (a) Schematic layout of the Graphene/MoS$_2$/Graphene devices (top). Cross section along the dotted line (bottom). (b) Optical image of a device. The green and black dashed lines indicate the edges of the graphene and the MoS$_2$, respectively. The spectra in (f), (g) and (h) were measured in regions with graphene, graphene overlapping with MoS$_2$ and MoS$_2$, respectively. (c) Photoluminescence mapping of MoS$_2$ at 1.87 eV. (d) Optical image of a Au/MoS$_2$/Au device. (e) Photoluminescence mapping of MoS$_2$ for the device in (d). (f) Raman spectrum of graphene (g) Raman and photoluminescence spectra of MoS$_2$ on graphene. (h) Raman and photoluminescence spectrum of MoS$_2$ before (black) and after (red) the patterning, transfer and alignment process described in supplementary material.

We characterized the MoS$_2$ devices with graphene and Au electrodes in ambient air and at room temperature. All of the samples were fabricated on doped silicon substrates covered with 300 nm of SiO$_2$, such that the substrate could be employed as a gate electrode. All of the devices were designed with a channel area of 10 μm × 60 μm. Figure 2 shows measurements of source-drain current as a function of the gate voltage (I$_{SD}$-V$_G$ curves) in linear and logarithmic scale for two of our devices, with graphene and gold contacts, respectively (other devices we measured and their characteristics are listed in the Supplementary Information). The *on/off* ratio and the mobility of the MoS$_2$ device with graphene electrodes, 1×10$^5$ and 0.48 cm$^2$V$^{-1}$s$^{-1}$ respectively, are both higher than those of the device with gold electrodes, the same parameters measuring 2 × 10$^3$ and 0.1 cm$^2$V$^{-1}$s$^{-1}$. As compared to the device with gold electrodes, the device with graphene electrodes showed a lower current in the *off* state and a higher current in the *on* state. This behavior is consistent with the different band alignments between graphene and MoS$_2$ and Au and MoS$_2$. Our CVD-grown graphene is hole-doped, with a Dirac point shifted to V$_G$ = 30 V (see the transfer characteristics of the graphene electrodes in the Supplementary Information), corresponding to a hole charge density of about 2× 10$^{12}$ cm$^{-2}$. This charge density lowers the Fermi energy with respect to intrinsic graphene by about 160 meV, leading to a larger work function, 4.66 eV. This value is still substantially smaller than the work function of Au, 5.1 eV[42-45]. As a result, the graphene/MoS$_2$ interface will have smaller Schottky barriers for electrons and larger Schottky barriers for holes than the Au/MoS$_2$ interface, leading to a larger electron current in the conduction band (*on* state) and smaller hole currents in the valence band (*off* state). We note that *on/off* ratios as high as 10$^8$ have been reported for devices based on exfoliated MoS$_2$, but such high ratios were usually obtained after annealing the samples to lower the contact resistance[40] and encapsulating them to remove impurities adsorbed on the MoS$_2$ channel. The devices shown in Fig. 2 are shown as-fabricated, with no annealing nor encapsulation, therefore their *on/off* ratios and mobilities could be increased by the means explained above.

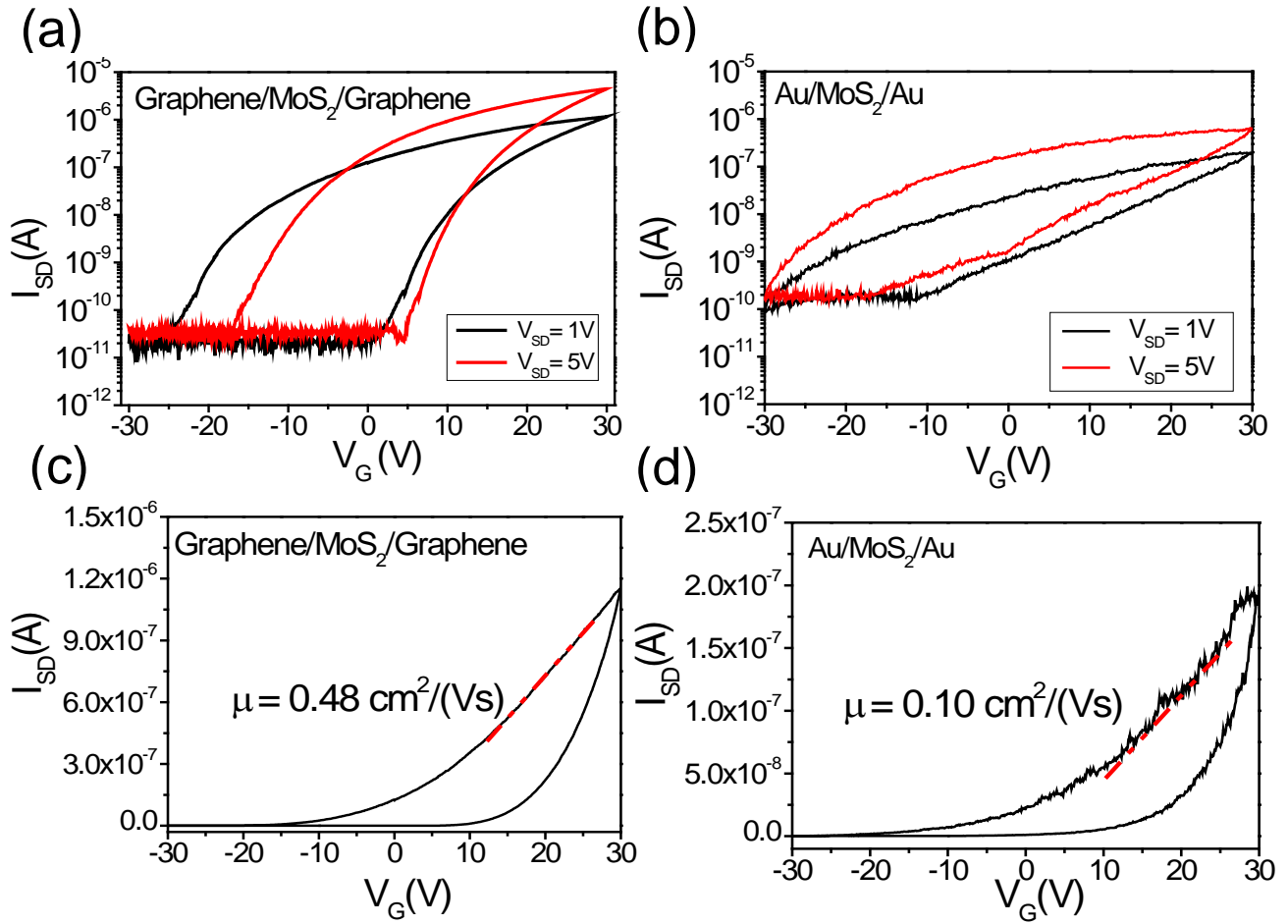

**Figure 2**. Source-drain current as a function of gate voltage for MoS$_2$ devices with graphene electrodes (a) and (c), and with gold electrodes, (b) and (d). The current is plotted in logarithmic scale in (a) and (b). For the curves in (c) and (d), V$_{SD}$ = 1 V.

Figures 2 and 3 show hysteresis in the I$_{SD}$-V$_G$ curves. This is likely related to impurities adsorbed on the MoS$_2$ that act as traps for electrons while the gate voltage is swept. Similar hysteresis has also been observed in devices made from exfoliated MoS$_2$ and can be reduced or eliminated with annealing and encapsulation of the devices.[26, 46]

Figures 3 (a) and (c) show the $I_{SD}$-$V_G$ curves of the graphene and Au contact devices exposed to 633-nm-wavelength light with different values of power density. Details of the optical set-up are in the Supplementary Information. The black curve is the $I_{SD}$-$V_G$ characteristic with the light off. The time-dependent response under illumination of 1 mWcm$^{-2}$, at different values of source-drain voltage and at a fixed gate voltage, $V_G$ = 30 V, is shown in Figures 3 (b) and (d). From these curves we calculate the photocurrent, $I_{PH}$, defined as the difference between the source-drain current under illumination (measured after the transient, when it reaches a stable value) and the source-drain dark current.

There are two main contributions to the photocurrent in MoS$_2$ photodetectors: photoconductivity and photogating.[41, 47] The photoconductive component is due to the increase in free charge carriers from photon absorption and the creation of electrons and holes. The photogating component is due to charge trapping from disorder and defects. The time dependent response can be fitted by the sum of two exponentials with different time constants, a shorter time constant varying from 10s to 25s and a longer one varying from 146s to 300s. Similar results for the time-dependence of the decay have been previously measured and attributed to different types of traps.[47, 48] A fraction of the photoexcited charge carriers fills these traps and changes the effective gate potential of the device, causing a threshold shift and a change in source-drain current. The photogating effect is most pronounced in the device with graphene contacts. As shown in the $I_{SD}$-$V_G$ curves in dark conditions in Figure 2, the device with graphene contacts has a larger source-drain current than the device with Au electrodes for positive gate voltages (larger current in the *on* state). Thus, a shift in the threshold voltage due to photogating will cause a larger photocurrent in the device with graphene electrodes.

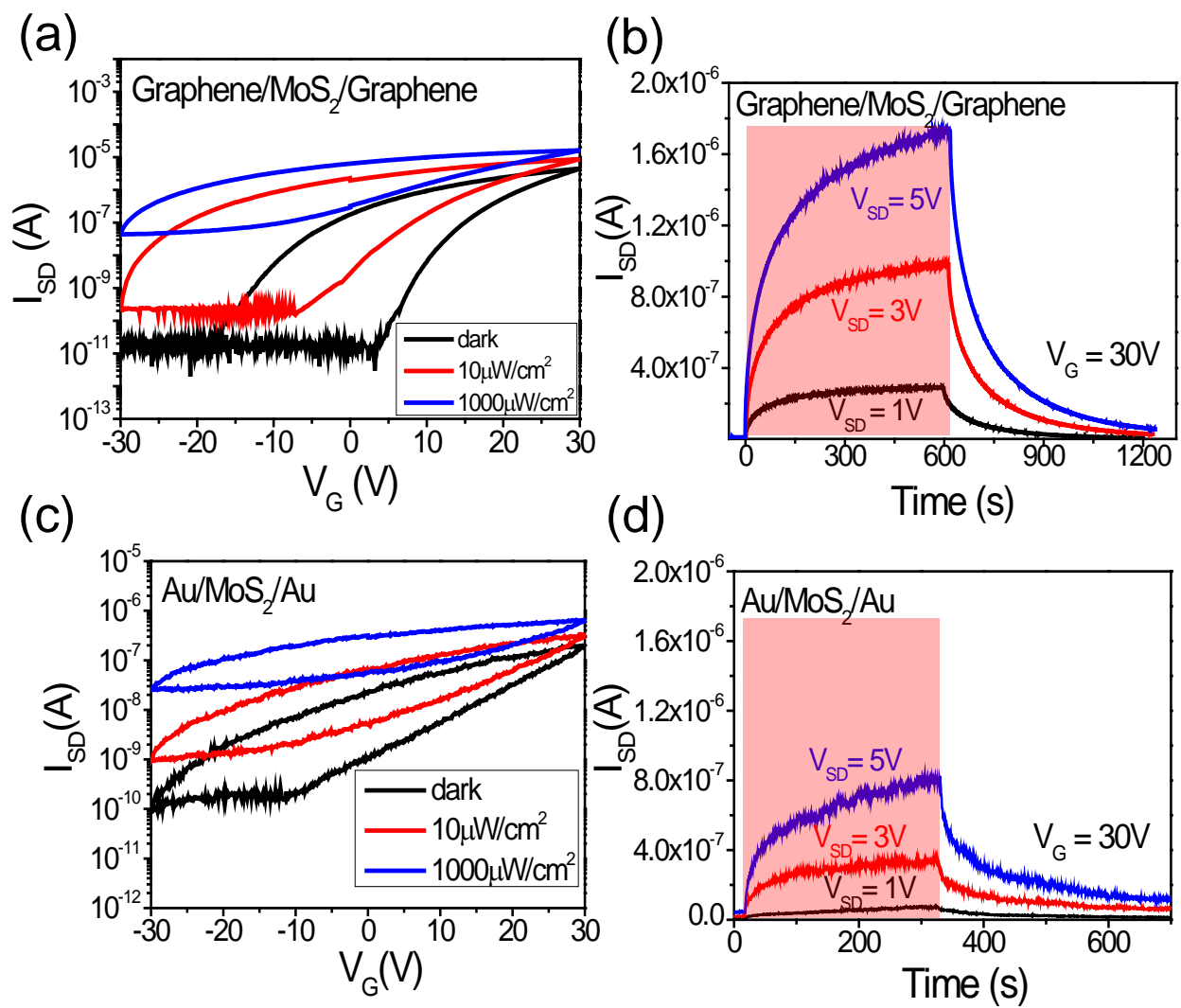

**Figure 3**. Gate dependence of the photoresponse for devices with graphene (a) and Au (c) contacts for different values of illumination power density from a source at 633 nm. The source-drain voltages are $V_{SD}= 5V$ for (a) and $V_{SD} = 1V$ for (c). The plots in (b) and (d) show the time-dependence of the photoresponse for three values of source-drain voltage and at a fixed illumination power density, 1000 μW/cm$^2$.

We evaluate the photoresponsivity for both devices, defined as the ratio between the photocurrent and the power P incident on the effective area of the device, $R = I_{PH}/P$. Since the power density from the source is known, we need to determine the effective area to calculate P. For the devices with Au electrodes, the sensing area is the area of the $MoS_2$ exposed to the light, i.e. the area of the $MoS_2$ between the Au electrodes. For devices with graphene electrodes, because graphene is highly transparent to visible light, we need to assess whether the sensing area extends from the $MoS_2$ in the gap between the graphene electrodes to the regions beyond that gap, where the $MoS_2$ and the graphene overlap. We measured the position-dependent photocurrent using a Raman spectrometer with a 532-nm source and a laser spot smaller than 1 μm. As shown in Figure 4 (a), the photocurrent decays quickly when the laser spot is moved outside the gap between the electrodes, therefore for both the graphene and Au contact devices, the area between the electrodes, about 10×60 μm$^2$, is a good estimate for our calculation of responsivities.

Figure 4 (b) shows the power dependence of the responsivity for the devices with graphene and gold contacts at two values of gate voltage, $V_G$ = +30V, -30V, corresponding to the field-effect transistors biased in the *on* and *off* state, respectively. At both values of gate voltage, the photoresponsivity decreases with increasing illumination power. This has also been observed in devices made from exfoliated $MoS_2$, where it has been attributed to long-lifetime traps becoming filled at low laser power, allowing short-lifetime traps to dominate the charge carrier dynamics at higher light intensities.[26, 40]

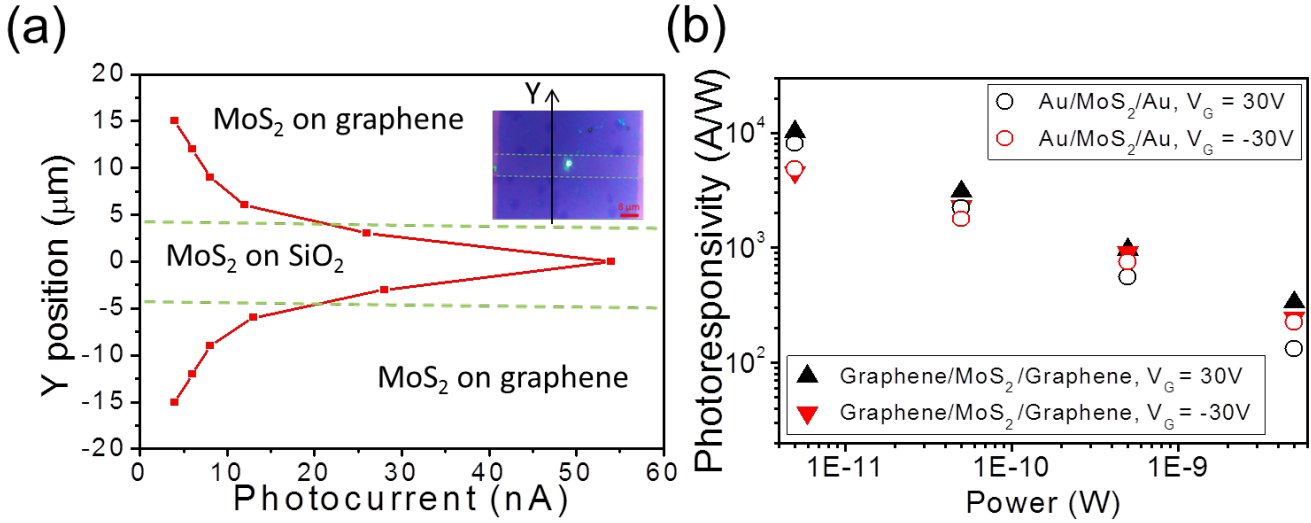

**Figure 4**. (a) Spatial dependence of the photocurrent for a Graphene/MoS$_2$/Graphene device at $V_G = 0$ and $V_{SD} = 1V$. The dashed green lines indicate the edges of the graphene electrodes. The inset is an optical image of the sample (the black line indicates the scanning direction of the laser and the bright spot is the laser). (b) Power dependence of the photoresponsivity for a Graphene/MoS$_2$/Graphene device and a Au/MoS$_2$/Au device at $V_{SD} = 5V$. The source wavelength is 532 nm for (a) and 633 nm for (b).

For both devices, the photoresponsivity is highest at the lowest illumination power and at values of gate voltage where the transistor is in the *on* state. The responsivity values are 10.24 kAW$^{-1}$ for the device with graphene electrodes and 8.02 kAW$^{-1}$ for the device with Au electrodes. Similarly high values of responsivities have also been reported for exfoliated MoS$_2$ devices.[26] A more significant figure of merit to compare the performance of different types of detectors for applications where high sensitivity is important is the specific detectivity, $D^* = R(AB)^{1/2}I_N^{-1}$, where $R$ is the photoresponsivity, $A$ is the detector area, $B$ the bandwidth and $I_N$ is the noise current[49, 50]. MoS$_2$ photodetectors show the highest detectivity at large negative gate biases where the source-drain dark current is the lowest and the noise is dominated by current shot noise. The highest reported value of shot-noise-limited detectivity for photodetectors based on $MoS_2$ is $8\times10^{13}$ Jones,[26] comparable to the values reported for other high-detectivity devices based multilayers of other layered semiconductors.[51-53] This is based on exfoliated $MoS_2$, encapsulated with $HfO_2$ and shot noise calculated from the dark current of the device in the *off* state, where the dark current is lowest.[26] We note that the shot-noise limited detectivity evaluates the performance limits of the devices, without taking in account 1/f noise due to the presence of defects and to the contacts. The work in Kufer and Konstantanos [26] shows that the measured noise indeed approaches the shot noise limit when the device is biased in the off state.

We calculated the shot-noise-limited detectivity of our devices for a bandwidth of 1 Hz, at $V_G$ = -30V, to be as high as $8.7 \times 10^{14}$ Jones for the graphene-contact device and $2.7 \times 10^{14}$ Jones for the Au-contact device. Both devices were measured in ambient conditions, with no encapsulation. To see the effect of encapsulation, we fabricated graphene/$MoS_2$/graphene devices where the $MoS_2$ encapsulated under the layer of PMMA that was used to transfer the $MoS_2$ to the device (see Supplementary Information). Figure 5 shows the $I_{SD}$-$V_G$ curves as a function of gate voltage for the encapsulated device. The PMMA encapsulation reduced the device hysteresis but did not eliminate it, indicating the presence of residual molecules that are likely adsorbed onto the surface of the $MoS_2$. The dark curve in Figure 5 (a) shows that the encapsulated device has a lower (higher) *on* (*off*) current than the graphene/$MoS_2$/graphene device without PMMA. This can be understood by noting that the graphene electrodes of the PMMA-covered device have a Dirac point that is shifted to about $V_G$ = 47 V, larger than the corresponding voltage for the exposed graphene, which indicates that they have stronger hole doping. We therefore estimate a larger work function for the PMMA-covered graphene, about 4.7 eV, leading to larger Schottky barriers for electrons in the conduction band and smaller Schottky barriers for holes in the valence band. The mobility of the PMMA-covered graphene-contact device is 0.59 cm$^2$V$^{-1}$s$^{-1}$, higher than that of the uncovered devices. This is consistent with results reported from other encapsulated devices, where the encapsulation also caused a higher mobility and a reduced hysteresis, likely due to a reduced concentration of adsorbed charged impurities.[26]

Figure 5 (a) and (b) also show the photoresponse of the encapsulated device and the power dependence of the photoresponsivity for values of gate voltage corresponding to the *on* and *off* state of the device. Similar to the other devices, the photoresponsivity is highest at lower values of incident power, but all the values are much higher than the uncovered devices and almost independent of gate voltage, yielding about $1.4\times10^5$ AW$^{-1}$ and $1.1\times10^5$ AW$^{-1}$ for $V_G$ = 10 V and $V_G$ = -30 V respectively. At negative gate voltage, where the dark current is lowest, we calculate a shot-noise limited detectivity D = $9 \times10^{15}$ Jones, about two orders of magnitude higher than the shot-noise limited detectivity obtained for devices based on exfoliated MoS$_2$. Although the characteristics varied among the devices we measured (see Table 1 in the Supplementary Information), all of them showed detectivities higher than $1 \times 10^{14}$ Jones.

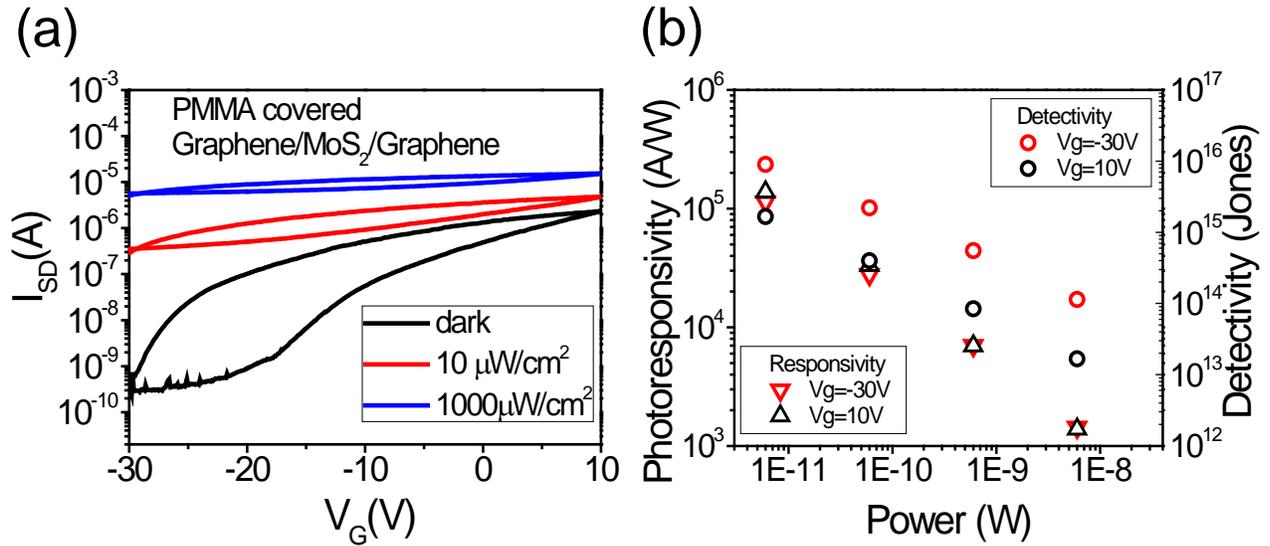

**Figure 5.** Photoresponse of a PMMA-covered Graphene/MoS$_2$/Graphene device to 633-nm light. (a) Source-drain current as a function of gate voltage for the device with light off (dark) and with two different values of illumination power. (b) Power dependence of the photoresponsivity at $V_G = 10$ V (black) and $V_G = -30$ V (red). $V_{SD} = 5$V for a) and b).

In conclusion, we fabricated photosensors based on CVD-grown monolayer MoS$_2$ by photolithography, using both gold and graphene electrodes. Our photosensors yield shot-noise limited detectivity a few orders of magnitude higher than photosensors from exfoliated MoS$_2$. The highest values of detectivity and responsivity were obtained from devices with graphene electrodes, paving the way to ultra-thin, highly sensitive photodetectors that are suitable for flexible substrates. The band alignment and the Schottky barriers between the graphene and MoS$_2$ can be tuned to optimize the device performance by varying the doping level of the graphene and the scalable fabrication process is promising for large-area photodetectors.

METHODS

*Synthesis of graphene*: The graphene was synthesized by CVD on copper films in ambient conditions. The copper film (Alfa Aesar 046365, 0.025mm, 99.8% purity) was cut from the roll, flattened, sonicated in a 5% nitric acid solution and deionized water, rinsed with IPA and dried with nitrogen gas flow. It was then placed on a quartz boat and pushed to the center of a horizontal 1-inch quartz tube furnace. The copper was annealed in Ar/ $H_2$ at 1030°C for 12 hours. The methane was added to start the graphene growth at 1045°C for 5 minutes. After the methane was turned off, the furnace was slowly cooled down to 500°C and then opened for fast cooling to room temperature, while flowing Ar and $H_2$ in the quartz tube.

*Synthesis of $MoS_2$*: Our growth substrates are Si chips coated with 300 nm $SiO_2$. They are cleaned with a hot piranha solution (1:3 $H_2O_2$:$H_2SO_4$) at 80°C for 10 minutes, rinsed with deionized water and dried with nitrogen. The PTAS solution was spun on the chips at 3000 rpm for 30s. The chips were then placed up-side down on the alumina boat containing $MoO_3$ powder (0.017g – 0.022g). This boat was put in the middle of the 2-inch quartz tube. A boat containing sulfur (0.045g-0.1g) was placed about 15cm-20cm upstream in the quartz tube, outside the furnace. The quartz tube was purged with 500 sccm Ar (99.999% purity) for 1 hour at 100 °C. Then the Ar flow rate was lowered to to 5 sccm and the furnace was heated to 650 °C at a rate of 15 °C/min. The temperature of sulfur was then raised to 180 °C and the $MoS_2$ was synthesized at 650 °C for 3 min under the atmospheric pressure. After growth, the samples were rapidly cooled down to room temperature by opening the furnace immediately and pulling the quartz tube out of the furnace. The Ar flow was increased to 500 sccm to remove the reactants.

*Photolithography process*: Two layers of PMMA 950 C2 (Microchem Inc) were spun on the surface of the substrates and then baked at 180°C on a hotplate to evaporate solvents. A layer of SU- 8 2002 (Microchem Inc) was subsequently spun on top of the PMMA and baked at 90°C for 10 min. The samples were exposed to UV light (i-line, 365nm) and then baked for another 10 minutes at 90°C. The samples were then immersed in the SU8 developer. The SU8 developer removed the SU8 that was not exposed to light, but it did not completely remove the PMMA underneath it. The residual PMMA was removed with oxygen plasma (at 30 mTorr, with gas flow 50sccm, forward power of 80W, for 3 minutes).

*Raman Spectroscopy*: Raman spectroscopy was performed with a Horiba LabRAM HR Evolution, with a 532 nm laser (Ventus 532 from Laser Quantum).

AUTHOR INFORMATION

**Corresponding Authors**

* Paola.Barbara@georgetown.edu; * ph523@georgetown.edu.

**Notes**

The authors declare no competing financial interests.

ACKNOWLEDGEMENTS

This work was supported by the by the US Office of Naval Research (award no. N000141310865 and N00014-16-1-2674) and the NSF (ECCS-1610953, REU, DMR-1358978, MRI, CHE-1429079 and DMR-0955625).

**Supporting Information:**

**Highly Sensitive Photodetectors with Graphene Contacts.**

*Peize Han\*[†], Luke St. Marie[†], Qing X. Wang[§], Nicholas Quirk[†], Abdel El Fatimy[†], Masahiro Ishigami[§] and Paola Barbara\*[†]*

[†]Department of Physics, Georgetown University, Washington, DC 20057, USA.

[§]Department of Physics and nanoscience Technology Center, University of Central Florida, Orlando, FL 32816, USA.

*$MoS_2$ transfer method:*

The steps for the $MoS_2$ transfer and alignment are shown in **Figure S1**. We start by coating the $MoS_2$ with PMMA. The edges of the samples are then placed in contact with thermal release tape (TRT) from the PMMA side, as shown in Figure 3. In the next step, we detach the $MoS_2$ from the $SiO_2$ surface, using the BHF wet etching described previously. Once the $MoS_2$ is detached from the substrate, while still being supported by the PMMA and the TRT, it is rinsed in DI water and then pulled out with tweezers. The PMMA and the TRT are then attached to a glass slide, as shown in Figure S1.

The $MoS_2$ is clearly visible through the glass slide in an optical microscope, therefore we can use the glass slide similarly to a regular mask in our mask aligner (Karl Suss MJB-3) to align the $MoS_2$ flakes or microribbons to the prepatterned graphene electrodes. Once the alignment is satisfactory, we place the $MoS_2$ in contact with the graphene. Next we remove the TRT by heating the whole structure at 130 ºC. As a final step, we remove the PMMA in acetone.

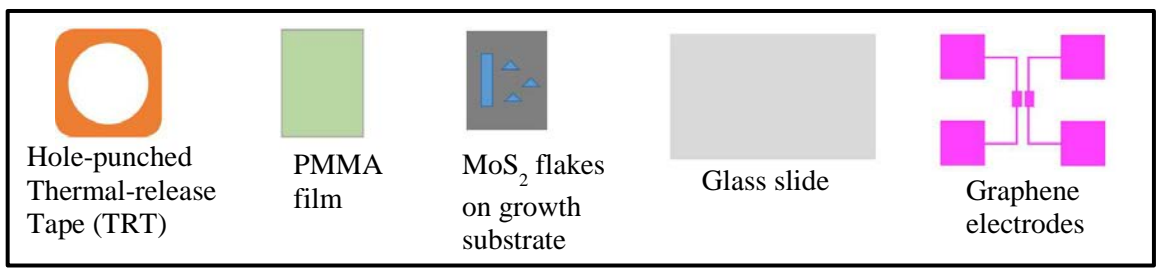

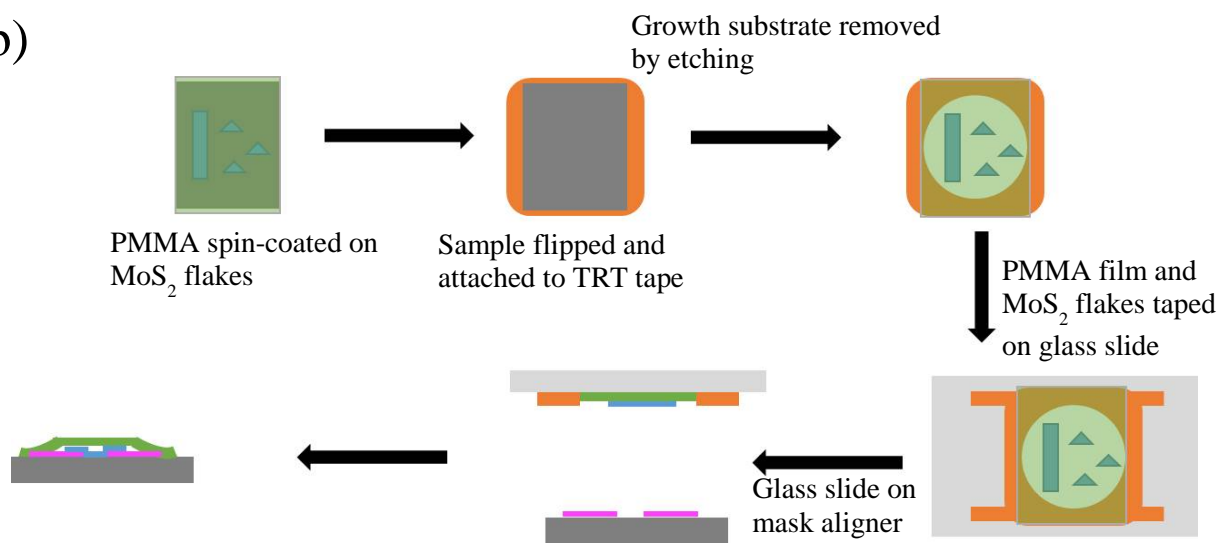

**Figure S1.** (a) Materials used for the transfer and alignment process. The steps are outlined in (b).

*Electrical characteristics of the graphene electrodes:* The resistance as a function of gate voltage curves for the graphene electrodes in the graphene/MoS$_2$/graphene devices discussed in the manuscript are shown in **Figure S2**. The graphene is hole-doped, with stronger doping for the PMMA-covered device (Figure S2 (b))

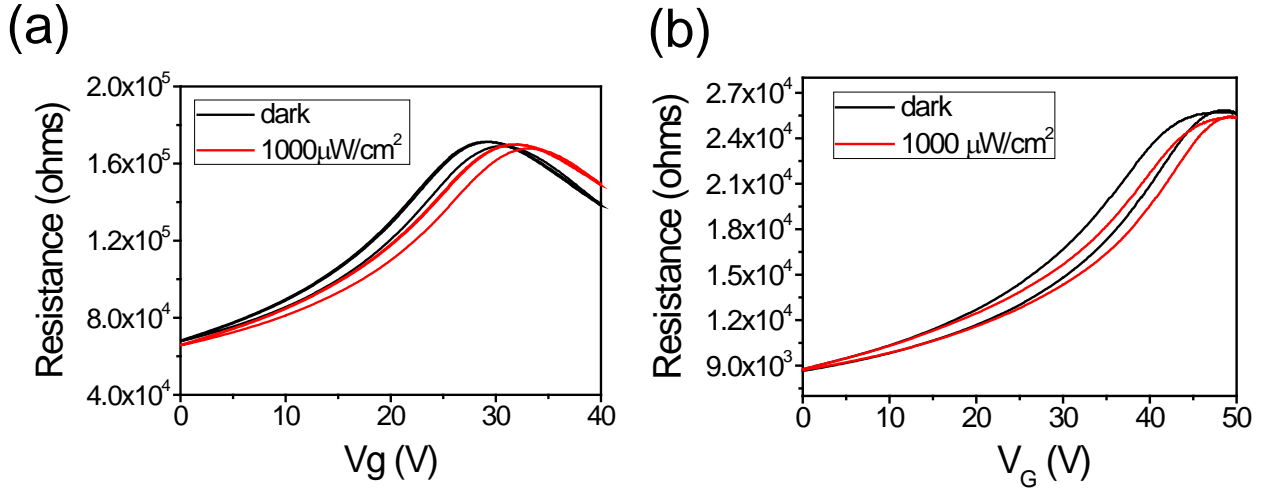

**Figure S2.** Transfer characteristics of the graphene electrodes for the exposed (a) and PMMA-covered (b) graphene/MoS2/graphene devices discussed in the manuscript.

*Photoresponse and characteristic parameters of all the samples measured:* The table below summarizes all the devices that we fabricated and characterized using the same geometry and the material growth and transfer methods described above.

All the devices were measured using the optical set-up sketched in **Figure S3**. The power density of the light was measured with a photometer from Industrial Fiber Optics (IF PM).

| device number | 1 | 2 | 3 * | 4 | 5 * | 6 | 7 * | 8 | 9 |
|---|---|---|---|---|---|---|---|---|---|
| layer number | 1 | 1 | 1 | 1 | 1 | 1 | 1 | 1 | 1 |
| channel length (μm) | 20 | 10 | 10 | 10 | 8 | 9 | 10 | 10 | 10 |
| channel width (μm) | 60 | 60 | 60 | 60 | 60 | 60 | 60 | 60 | 60 |
| electrode material | Au | Au | Au | Au | graphene | graphene | graphene | graphene | graphene |
| PMMA encapsulation | no | no | no | no | no | no | yes | yes | no |
| Vsd (V) | 1 | 1 | 1 | 5 | 1 | 4 | 5 | 5 | 5 |
| Vg (V) | ±30 | 10 | ±30 | ±30 | ±30 | gate leak | ±30 | ±30 | ±30 |
| mobility (cm^2/V·s) | 0.1 | 0.13 | 0.1 | 0.22 | 0.48 | NA | 0.59 | 0.05 | 0.14 |
| on-off ratio | 5.4E+03 | 3.7E+03 | 2.1E+03 | 3.3E+03 | 1.2E+05 | NA | 4.6E+03 | 1.9E+03 | 3.4E+03 |
| responsivity (kA/W) | 0.035 @Vsd 5V, Vg 0V for 50μW/cm^2 | NA | 4.8 @Vg -30V, Vsd 5V for 1μW/cm^2 | 1.9 @Vg -30V, Vsd 5V for 1μW/cm^2 | 4.5 @Vg -30V, Vsd 5V for 1μW/cm^2 | 0.81 @Vg 0V, Vsd 4V for 50μW/cm^2 | 114 @Vg -30V, Vsd 5V for 1μW/cm^2 | 11.5 @Vg -30V, Vsd 5V for 1μW/cm^2 | 3.2 @Vg -30V, Vsd 5V for 1μW/cm^2 |
| detectivity (Jones), Vg= -30V | NA | NA | 2.7E+14 | 3.7E+14 | 8.8E+14 | NA | 9.0E+15 | 1.6E+15 | 6.3E+14 |

**Table 1.** Summary of all the devices tested and their properties. The devices discussed in the manuscript are those labelled with an asterisk.

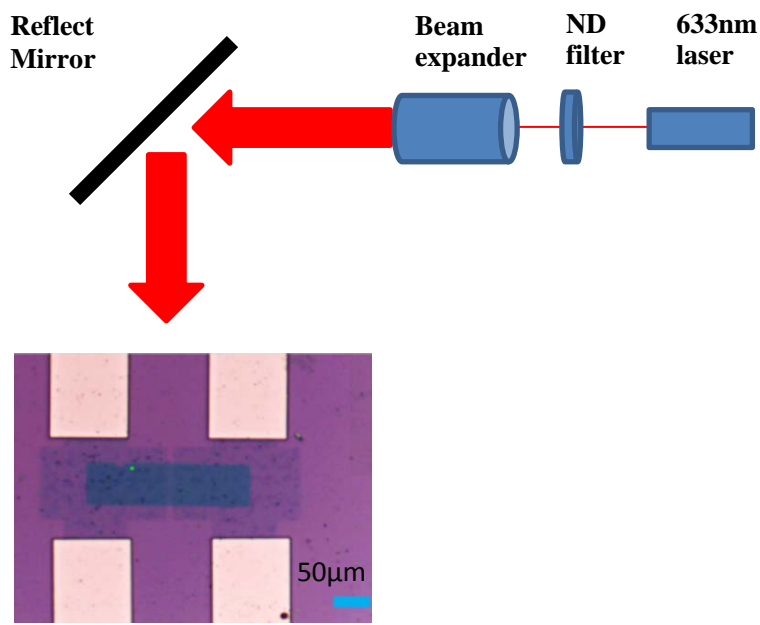

**Figure S3.** Sketch of the optical set-up. For all the experiments, the beam size was about 1 cm in diameter.

*Optical characterization of the Graphene/MoS$_2$/Graphene devices:* **Figure S4** shows the spectrum of MoS$_2$ measured in the gap between the graphene electrodes and the spectrum measured in the area where the graphene and the MoS$_2$ overlap. The photoluminescence of MoS$_2$ is reduced in the overlap region, due to the charge transfer between MoS$_2$ and graphene[1, 2]

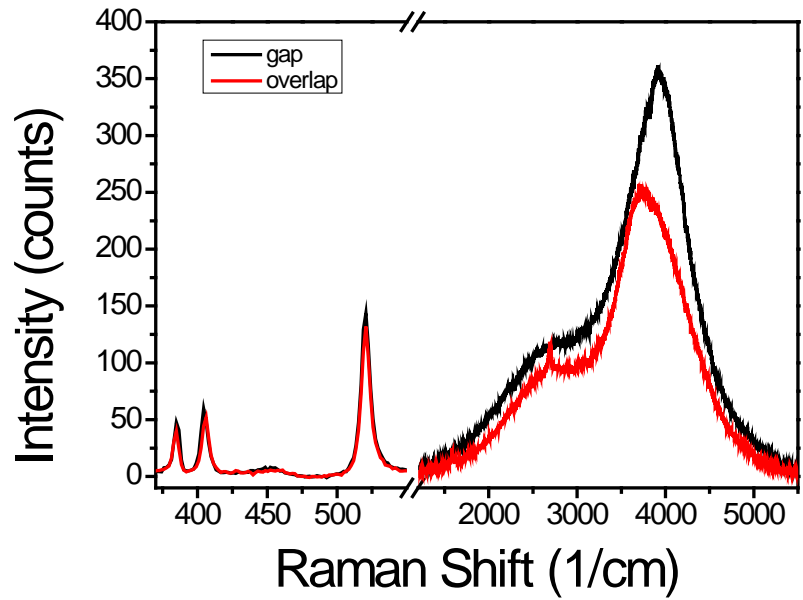

**Figure S4.** The Raman and photoluminescence spectrum of MoS$_2$ in the gap (black) and MoS$_2$ on graphene (red).

*Annealing tests:* Most of measurements reported in the manuscript were performed in air, soon after the sample fabrication, without any annealing. After a few months, we tested the samples again and found that the mobility was decreased from the values reported in table 1 and the hysteresis was increased. This is expected, since the mobility decreases with time for devices that are not sealed, due to adsorption of molecules on the $MoS_2$.[3]

We tested two annealing procedures. In the first procedure the samples were annealed in high vacuum ($5\times10^{-7}$ Torr), at 100 °C for 18 hours.[4] In the second procedure the samples were annealed in ambient pressure with forming gas (100 sccm Ar and 10 sccm $H_2$) at 200 °C for two hours.[6] Unless otherwise noted, following each annealing, a thin layer of PMMA was spin-coated and baked at 180 °C to provide a temporary passivation of the device.[5] The PMMA was removed with acetone before subsequent annealing. Below are the results from the two annealing processes on samples with Au contacts and with graphene contacts (samples #3 and #9 in Table 1). Sample #9 was first annealed in vacuum, tested and then annealed with forming gas. Sample #3 was first annealed in forming gas (no PMMA coating after), tested and then annealed in vacuum (with PMMA coating after).

**Annealing in high vacuum:** The conductance vs. gate voltage and the photoresponse for samples #3 and #9 before and after annealing are shown in Figures S5 and S6. (Note that even though sample #3 had been previously annealed in forming gas, as shown in the next section, its mobility decreased quickly after the first annealing due to ambient exposure to air because it was not coated with PMMA.) In both cases the *on* state conductance and the mobility increased. The hysteresis was substantially reduced for the sample with graphene contacts.

Tables 2 and 3 show the characteristic parameters of the photoresponse before and after annealing, for samples #3 and #9, respectively. The detectivity values are similar to the values measured a few months ago after the sample fabrication.

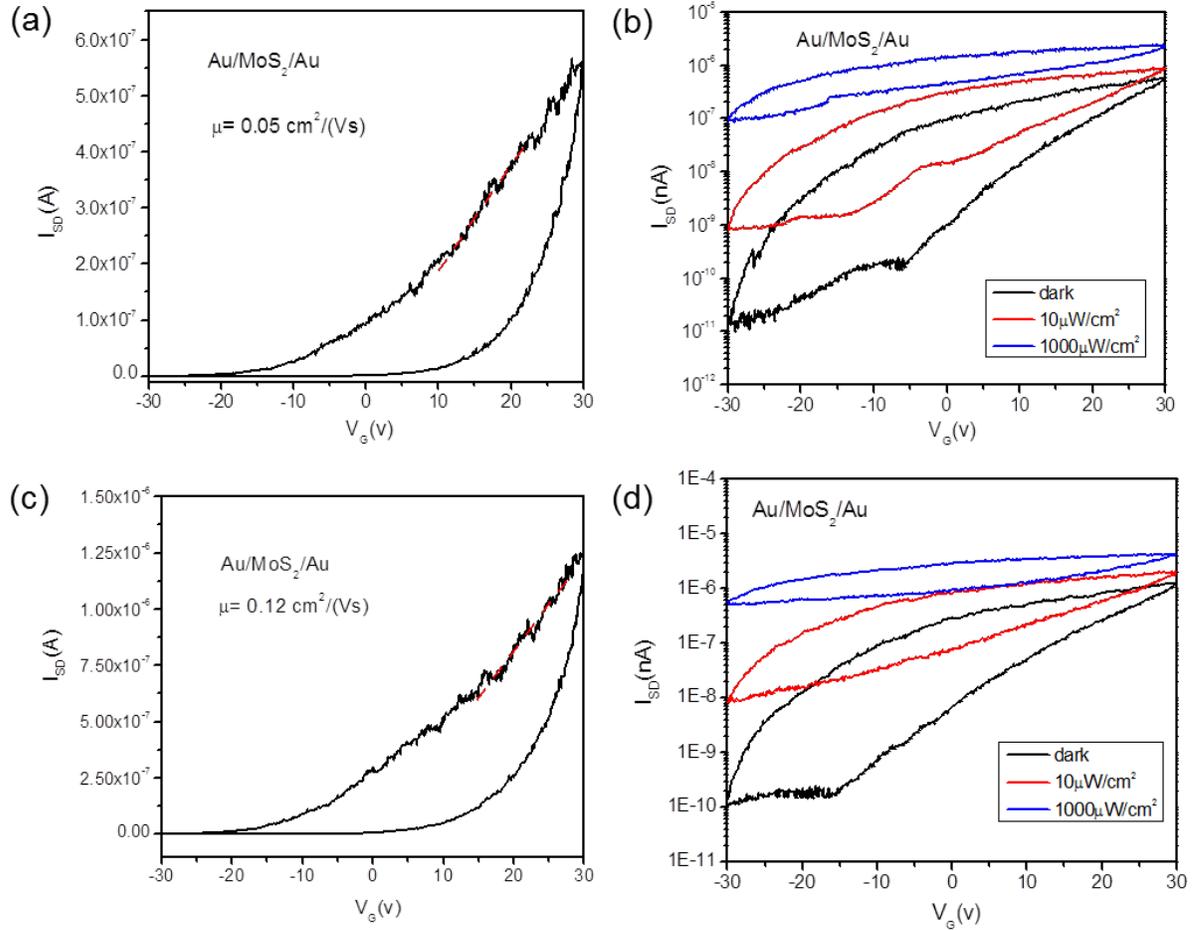

**Figure S5.** Source-drain current as a function of gate voltage for the Au/MoS$_2$/Au device (#3 in Table 1) before (a) and after (c) vacuum annealing with $V_{SD}$ = 5V. Photoresponse before (b) and after (d) vacuum annealing for different values of illumination power density from a source at 633nm.

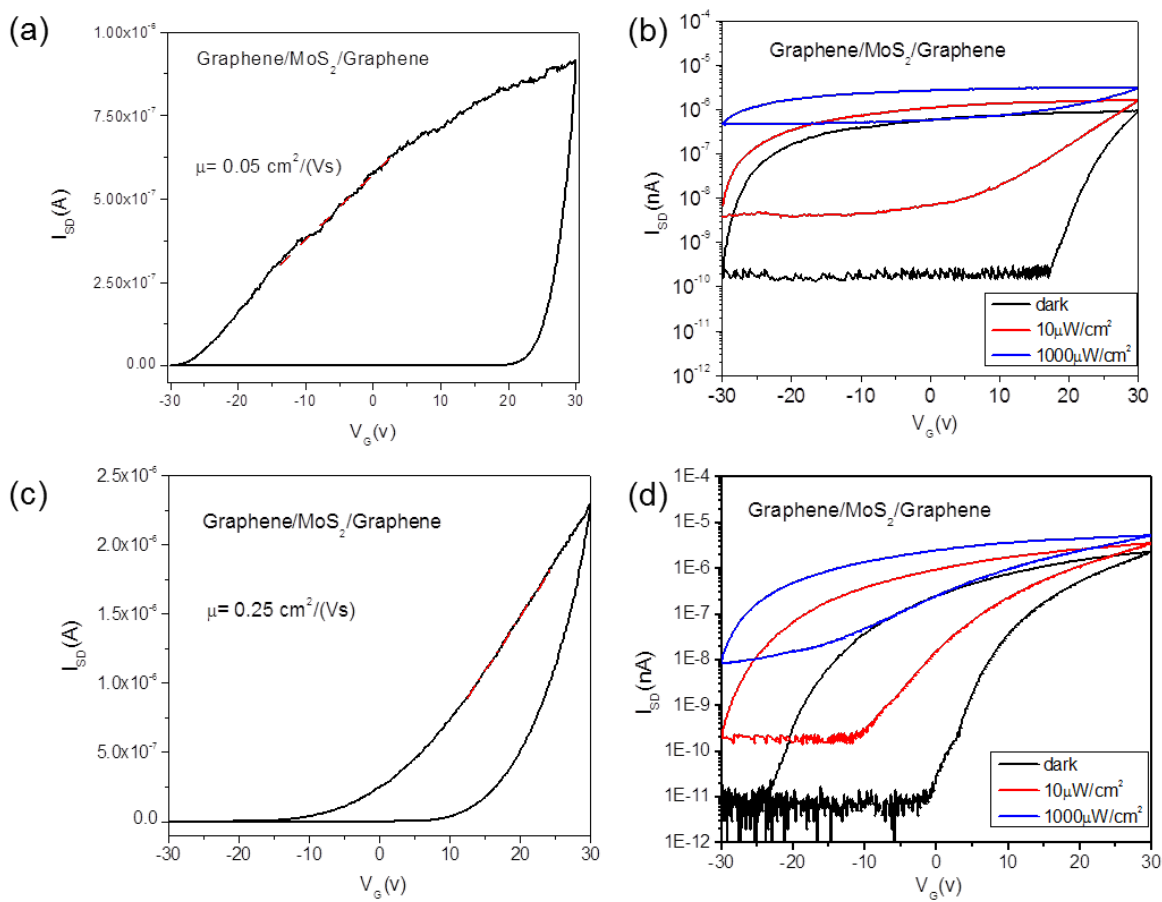

**Figure S6.** Source-drain current as a function of gate voltage for the graphene/MoS$_2$/Graphene device (#9 in Table 1) before (a) and after (c) vacuum annealing with V$_{SD}$ = 5V. Photoresponse before (b) and after (d) vacuum annealing for different values of illumination power density from a source at 633nm.

We also tested the effect of annealing on the response time of the device. The data are reported in Table 4. Similarly to previous work[3, 7], the rise and decay time of the time-dependent curves (see Figure 3 (b) and (d) in the manuscript) can be fitted by the sum of two exponential functions with different decay times, t1 and t2, due to the presence of different types of traps.[3, 7]

| Illumination density (μW/cm²) | Dark current (nA) | Photocurrent (nA) | Illumination (W) | Photoresponsivity (10^3 A/W) | Detectivity (jones) |
|---|---|---|---|---|---|
| Before vacuum anneal | | | | | |
| Au (#3) @+30V | | | | | |
| 10 | 97 | 73 | 6E-11 | 1.2 | 1.7E+13 |
| 1000 | 130 | 640 | 6E-09 | 0.1 | 1.3E+12 |
| After vacuum anneal | | | | | |
| Au (#3) @+30V | | | | | |
| 1 | 60 | 47.5 | 6E-12 | 3.1 | 5.5E+13 |
| 10 | 60.6 | 109.7 | 6E-11 | 1.4 | 2.4E+13 |
| 100 | 61 | 389 | 6E-10 | 0.4 | 7.8E+12 |
| 1000 | 70 | 893 | 6E-09 | 0.1 | 1.7E+12 |
| Au (#3) @-30V | | | | | |
| 10 | 2.5 | 18.2 | 6E-11 | 0.3 | 2.6E+13 |
| 100 | 9 | 100 | 6E-10 | 0.2 | 7.6E+12 |
| 1000 | 16 | 826 | 6E-09 | 0.1 | 4.7E+12 |

**Table 2.** Photoresponsivity and detectivity for the Au/MoS$_2$/Au device before and after the vacuum annealing.

| Illumination density (μW/cm²) | dark current (nA) | Photocurrent (nA) | Illumination (W) | Photoresponsivity (10^3 A/W) | Detectivity (jones) |
|---|---|---|---|---|---|
| Before vacuum anneal | | | | | |
| Gr (#9) @+30V | | | | | |
| 1 | 72.3 | 30.5 | 6E-12 | 5.1 | 8.2E+13 |
| 10 | 79.8 | 81.0 | 6E-11 | 1.4 | 2.1E+13 |
| 100 | 82.3 | 403.7 | 6E-10 | 0.7 | 1.0E+13 |
| 1000 | 97.6 | 1114.4 | 6E-09 | 0.2 | 2.6E+12 |
| Gr (#9) @-30V | | | | | |
| 1 | 24.7 | 32.7 | 6E-12 | 5.5 | 1.5E+14 |
| 10 | 42 | 100 | 6E-11 | 1.7 | 3.5E+13 |
| 100 | 54.8 | 319.2 | 6E-10 | 0.5 | 9.8E+12 |
| 1000 | 83 | 983 | 6E-09 | 0.2 | 2.5E+12 |
| After vacuum anneal | | | | | |
| Gr (#9) @+30V | | | | | |
| 1 | 1 | 18.5 | 6.0E-12 | 3.1 | 4.2E+14 |
| 10 | 2 | 83.2 | 6.0E-11 | 1.4 | 1.3E+14 |
| 100 | 4 | 268 | 6.0E-10 | 0.4 | 3.1E+13 |
| 1000 | 4.2 | 615.8 | 6.0E-09 | 0.1 | 6.9E+12 |
| Gr (#9) @-30V | | | | | |
| 1 | 0.05 | 8.8 | 6.0E-12 | 1.5 | 8.9E+14 |
| 10 | 0.7 | 49.3 | 6.0E-11 | 0.8 | 1.3E+14 |
| 100 | 1.6 | 185.2 | 6.0E-10 | 0.3 | 3.4E+13 |
| 1000 | 4.2 | 535.8 | 6.0E-09 | 0.1 | 5.9E+12 |

**Table 3.** Photoresponsivity and detectivity for the graphene/MoS$_2$/graphene device before and after the vacuum annealing

We noticed that after annealing the shorter decay time decreased, while the shorter raise time increased. In all cases, time constants are similar to the time constants reported in previous work for photodetectors operating in ambient conditions[3, 7]. The effect of annealing on the response time of these devices is still under investigation.

| device | vacuum anneal | rise or decay | Vg (V) | Vsd (V) | t1 (s) | t2 (s) |
|---|---|---|---|---|---|---|
| #3 | before | rise | 30 | 5 | 7.1 | 95.9 |
| #3 | before | decay | 30 | 5 | 55.9 | 715.5 |
| #3 | after | rise | -30 | 5 | 19.3 | 112.2 |
| #3 | after | decay | -30 | 5 | 18.6 | 234.2 |
| #3 | after | rise | 30 | 5 | 14.2 | 94.5 |
| #3 | after | decay | 30 | 5 | 19.6 | 193.6 |
| #9 | before | rise | -30 | 5 | 16.8 | 116.5 |
| #9 | before | decay | -30 | 5 | 24.2 | 117.6 |
| #9 | before | rise | 30 | 5 | 11.7 | 89.1 |
| #9 | before | decay | 30 | 5 | 26.2 | 125.1 |
| #9 | after | rise | -30 | 5 | 27.9 | 206.1 |
| #9 | after | decay | -30 | 5 | 18.6 | 171.1 |
| #9 | after | rise | 30 | 5 | 48.3 | 396.3 |
| #9 | after | decay | 30 | 5 | 22 | 227.7 |

**Table 4.** Characteristic times of the photoresponse for the Au/MoS$_2$/Au device (#3) and the graphene/MoS$_2$/graphene device (#9) before and after the vacuum annealing.

**Annealing in forming gas:** The conductance vs. gate voltage curves and the photoresponse of devices #3 and #9 before and after the annealing in forming gas are shown in Figures S7 and S8. In both cases, the *on* current and the mobility increased, with a much larger increase for the device with Au contacts. The effect of the annealing on the photoresponse are summarized in Tables 5 and 6. Even for the sample with Au contacts, the detectivity stayed similar to the values measured for the device right after fabrication.

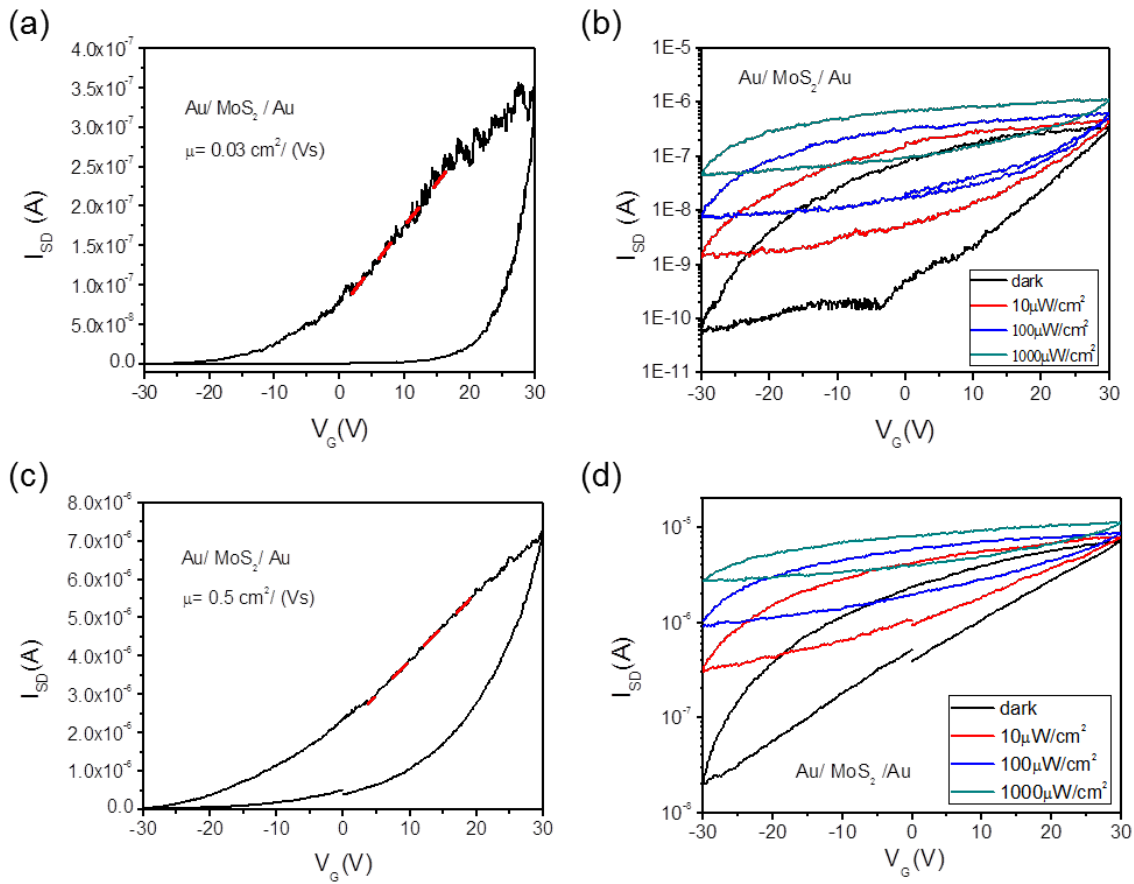

**Figure S7.** Source-drain current as a function of gate voltage for the Au/MoS$_2$/Au device (#3 in Table 1) before (a) and after (c) annealing in forming gas with $V_{SD} = 5$V. Photoresponse before (b) and after (d) annealing in forming gas for different values of illumination power density from a source at 633nm

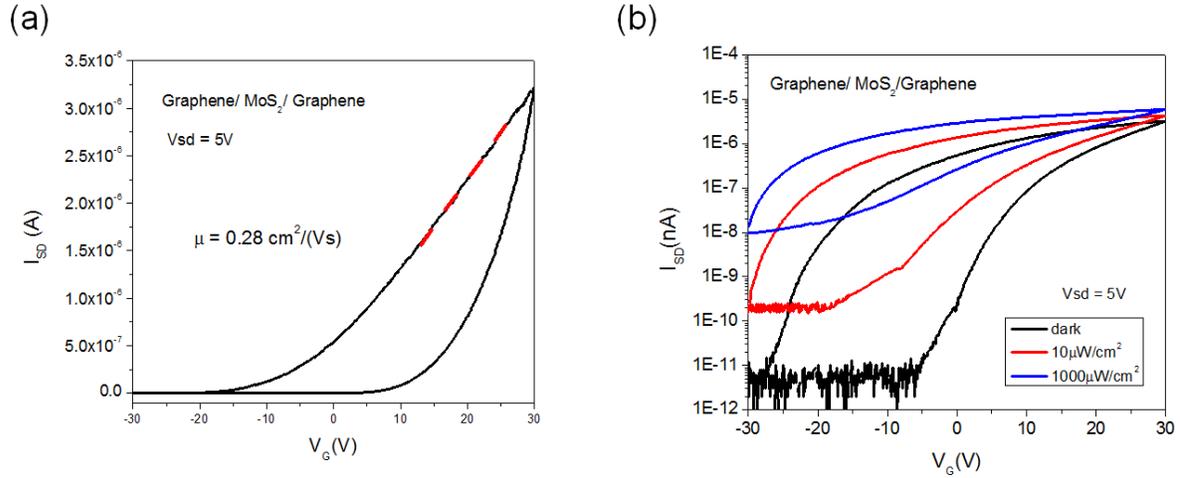

**Figure S8.** Source-drain current and photoresponse as a function of gate voltage for the graphene/MoS$_2$/Graphene device (#9 in Table 1) after annealing in forming gas, with V$_{SD}$ = 5V. The characteristic before annealing were unchanged from those shown in Figures S6 (c) and (d) (the sample was protected with PMMA after the first vacuum annealing and the characteristics did not degrade).

| Illumination density (μW/cm$^2$) | dark current (nA) | Photocurrent (nA) | Illumination (W) | Photoresponsivity (10^3 A/W) | Detectivity (jones) |
|---|---|---|---|---|---|
| Before anneal | | | | | |
| Vg -30 (#3) | | | | | |
| 1 | 6 | 29 | 6E-12 | 4.8 | 2.7E+14 |
| After anneal | | | | | |
| Vg 30 (#3) | | | | | |
| 1 | 70 | 100 | 6E-12 | 16.7 | 2.7E+14 |
| 10 | 75 | 305 | 6E-11 | 5.1 | 8.0E+13 |
| 100 | 110 | 790 | 6E-10 | 1.3 | 1.7E+13 |
| 1000 | 160 | 1440 | 6E-09 | 0.2 | 2.6E+12 |
| Vg -30 (#3) | | | | | |
| 1 | 14 | 41 | 6E-12 | 6.8 | 2.5E+14 |
| 10 | 28 | 142 | 6E-11 | 2.4 | 6.1E+13 |
| 100 | 55 | 414 | 6E-10 | 0.7 | 1.3E+13 |
| 1000 | 70 | 1130 | 6E-09 | 0.2 | 3.1E+12 |

**Table 5.** Photoresponsivity and detectivity for the Au/MoS$_2$/Au device before and after the annealing in forming gas.

| Illumination density (μW/cm$^2$) | dark current (nA) | Photocurrent (nA) | Illumination (W) | Photoresponsivity (10^3 A/W) | Detectivity (jones) |
|---|---|---|---|---|---|
| After anneal | | | | | |
| Vg 30 (#9) | | | | | |
| 1 | 38 | 84 | 6E-12 | 7.7 | 1.7E+14 |
| 10 | 42 | 143 | 6E-11 | 1.7 | 3.6E+13 |
| 100 | 43 | 322 | 6E-10 | 0.5 | 9.7E+12 |
| 1000 | 42 | 724 | 6E-09 | 0.1 | 2.4E+12 |
| Vg -30 (#9) | | | | | |
| 1 | 1 | 11 | 6E-12 | 1.7 | 3.1E+14 |
| 10 | 2 | 62 | 6E-11 | 1.0 | 1.1E+14 |
| 100 | 3 | 217 | 6E-10 | 0.4 | 2.6E+13 |
| 1000 | 5 | 567 | 6E-09 | 0.1 | 5.7E+12 |

**Table 6.** Photoresponsivity and detectivity for the graphene/MoS$_2$/graphene device after the annealing in forming gas.

The effect of annealing in forming gas on the response time is summarized in Table 7. Only the characteristic times for the device with Au contact were substantially shortened.

| device | before or after | rise or decay | V$_G$ (V) | V$_{SD}$ (V) | t1 (s) | t2 (s) |
|---|---|---|---|---|---|---|
| #3 | before | rise | 30 | 5 | 11 | 212.8 |
| #3 | before | decay | 30 | 5 | 49.4 | 529.8 |
| #3 | after | rise | 30 | 5 | 7.1 | 83.3 |
| #3 | after | decay | 30 | 5 | 31.8 | 298.2 |
| #9 | after | rise | -30 | 5 | 25.7 | 302.9 |
| #9 | after | decay | -30 | 5 | 41.4 | 414.5 |
| #9 | after | rise | 30 | 5 | 48.2 | 484.4 |
| #9 | after | decay | 30 | 5 | 67.7 | 531.2 |

**Table 7.** Characteristic times of the photoresponse for the Au/MoS$_2$/Au device (#3) and the graphene/MoS$_2$/graphene device (#9) before and/or after annealing in forming gas.